\documentclass[11pt]{article}
\usepackage{amsfonts,amsmath}
\def\bbt{\bibitem}
\def\be{\begin{equation}}
\def\en{\end{equation}}
\def\ber{\begin{eqnarray}}
\def\enr{\end{eqnarray}}
\def\nmb{ \nonumber\\}
\def\d{\partial}

\def\ov{\over }
\def\tld{\tilde}

\def\sgm{\sigma}

\def\al{\alpha}

\def\gm{\gamma}
\def\Gm{\Gamma}
\def\im{\imath}

\def\lm{\lambda}
\def\Lm{\Lambda}
\def\Om{\Omega}

\def\et{\eta}

\def\ups{\upsilon}
\def\Ups{\Upsilon}
\def\dlt{\delta}
\def\Dl{\Delta}
\def\kp{\kappa}

\def\bh{{\bf h}}
\def\bp{{\bf p}}

\def\bt{{\bf t}}

\def\bv{{\bf v}}
\def\bd{{\bf d}}

\def\be{{\bf e}}
\def\bmu{{\boldsymbol \mu}}
\def\bLm{{\boldsymbol \Lm}}

\def\blm{{\boldsymbol \lambda}}
\def\bphi{{\boldsymbol \phi}}

\begin{document}
\vskip 2 true cm
\centerline{\bf FREE-FIELD REPRESENTATION OF PERMUTATION BRANES}
\centerline{\bf IN GEPNER MODELS.}
\vskip 1.5 true cm
\centerline{\bf S. E. Parkhomenko}
\centerline{Landau Institute for Theoretical Physics}
\centerline{142432 Chernogolovka, Russia}
\vskip 0.5 true cm
\centerline{spark@itp.ac.ru}
\vskip 1 true cm
\centerline{\bf Abstract}
\vskip 0.5 true cm

 We consider free-field realization of Gepner models basing on
free-field realization of N=2 superconformal minimal models. 
Using this realization we analyse A/B-type boundary conditions
starting from the ansatz when left-moving and right-moving
free-fields degrees of freedom are glued at the boundary by an arbitrary 
constant matrix. It is shown that the only boundary conditions consistent with
the singular vectors structure of unitary minimal models representations
are given by permutation matrices and give thereby 
explicit free-field construction of permutation branes of Recknagel. 

\vskip 10pt

"{\it PACS: 11.25Hf; 11.25 Pm.}"

{\it Keywords: Strings, $D$-branes,
Conformal Field Theory.}

\smallskip
\vskip 10pt
\centerline{\bf0. Introduction}
\vskip 10pt

 The investigation of D-branes on Calabi-Yau manifolds on string scales is 
interesting and important problem. There is a significant progress in this
direction achieved mainly due to the intensive study of D-branes at Gepner points
of Calabi-Yau moduli space initiated by Recknagel and Shomerus ~\cite{ReS}.

 Because of Gepner models are defined by puerly algebraic
construction ~\cite{Gep}, ~\cite{EOTY} it is natural that the symmetry 
preserving boundary states (D-branes) in these models can be described by 
algebraic objects also ~\cite{ReS}-~\cite{Reper}. Thus the question of their 
geometric interpretation appears to be nontrivial and interesting.
The considerable
progress in the understanding of the geometry of $D$-branes in Gepner models
has been achieved recently in ~\cite{BDLR}-~\cite{Tom}.
The main idea developed in these papers is to relate the intersection index of 
boundary states ~\cite{DFi} with the bilinear form of the $K$-theory classes of 
bundles on the large volume CY manifold and use this relation to associate 
the $K$-theory classes to the boundary states. 

 The natural question which appears in this concern but hard to answer is 
if one can find direct CFT description of geometry of D-branes in Gepner models
instead of interpolation of large-volume topological dates of bundles into 
the Gepner point? Trying to find the direct description (as well as to develop integral 
representation for the boundary correlation functions) the free-field construction 
of D-branes in Gepner models has been developed in ~\cite{SP}. It was shown there 
that the free-field representations of
the open string spectrum between the Recknagel-Shomerus boundary states 
can be described in terms of representations of Malikov, Schechtman and Vaintrob 
chiral de Rham complex ~\cite{MSV} on the Landau-Ginzburg orbifold. 
The chiral de Rham complex is string generalization of the usual de Rham complex 
and is a sheaf of vertex algebras ~\cite{MSV}-~\cite{GobM}. Hence it is a 
geometric object and this property has been used in ~\cite{SP} to interprate geometrically 
the boundary states in Gepner models (constructed in puerly algebraic terms) 
as a fractional branes on Landau-Ginzburg orbifolds. 
This suggests that chiral de Rham complex might be natural and
efficient object for the description of D-brane geometry at string scales.

 Having this in mind we try in this note to extend the free-field representation 
of ~\cite{SP}, ~\cite{SPmm} for the case of permutation branes ~\cite{Reper}.
Our aim here is to analyse and represent the free-field construction of permutation
branes leaving the important question of study and comparing of the free-field 
geometry of D-branes to the results of ~\cite{BrGa}, ~\cite{ERR} for the future.

 In section 1 we briefly review free-field construction of irreducible 
representations in $N=2$ minimal models developed by Feigin and Semikhatov. 
In section 2 we schematically consider the free-field
realization of Gepner models. In section 3 we investigate A and B-type gluing conditions
in terms of the free-fields. We start from the ansatz where the 
left-moving and right-moving free-field degrees
of freedom are glued at the boundary by constant arbitrary matrix and analyse 
A and B-type boundary conditions in terms of free-fields.
The section 4 is a main part of the paper. We analyse the consistency of the boundary 
conditions with the singular vectors structure 
of minimal models (butterfly resolution) and show that only 
permutation matrices survive giving thereby
the free-field representation of permutation Ishibashi states. 
In section 5 we use the Recknagel solution ~\cite{Reper} of Cardy's constraints 
as well as orbifold construction to obtain free-field realization of permutations 
branes in Gepner models.

\vskip 10pt
\centerline{\bf 1. Free-field realization of $N=2$ minimal models}
\centerline{\bf irreducible representations.}
\vskip 10pt

 In this section we briefly discuss free-field construction of
Feigin and Semikhatov ~\cite{FeS} of the irreducible modules
in $N=2$ superconformal minimal models. Free-field approach to $N=2$ minimal models
considered also in \cite{Ohta}-~\cite{Wit}.

\leftline{\bf 1.1. Free-field representations of $N=2$ super-Virasoro
algebra.}

  We introduce (in the left-moving sector) the free bosonic fields
$X(z), X^{*}(z)$ and free fermionic fields $\psi(z), \psi^{*}(z)$,
so that its OPE's are given by
\ber
X^{*}(z_{1})X(z_{2})=\ln(z_{12})+reg.,\nmb
\psi^{*}(z_{1})\psi(z_{2})=z_{12}^{-1}+reg.,
\label{1.ope}
\enr
where $z_{12}=z_{1}-z_{2}$. Then for an arbitrary number
$\mu$ the currents of $N=2$ super-Virasoro
algebra are given by
\ber
G^{+}(z)=\psi^{*}(z)\d X(z) -{1\ov \mu} \d \psi^{*}(z), \
G^{-}(z)=\psi(z) \d X^{*}(z)-\d \psi(z), \nmb
J(z)=\psi^{*}(z)\psi(z)+{1\ov \mu}\d X^{*}(z)-\d X(z), \nmb
T(z)=\d X(z)\d X^{*}(z)+
{1\ov 2}(\d \psi^{*}(z)\psi(z)-\psi^{*}(z)\d \psi(z))-\nmb
{1\ov 2}(\d^{2} X(z)+{1\ov \mu}\d^{2} X^{*}(z)),
\label{1.min}
\enr
and the central charge is
\ber
c=3(1-{2\ov \mu}).
\label{1.cent}
\enr

 As usual, the fermions $\psi(z),\psi^{*}{z}, G^{\pm}(z)$ in NS (R) sector 
are expanded into half-integer (integer) modes. The bosons 
$X(z),X^{*}(z),J(z),T(z)$ are expanded into integer
modes in both sectors. 

 In NS sector $N=2$ Virasoro superalgebra is acting naturally in Fock module
$F_{p,p^{*}}$ generated by the fermionic
operators $\psi^{*}[r]$, $\psi[r]$, $r<{1\ov2}$, and bosonic operators
$X^{*}[n]$, $X[n]$, $n<0$ from the vacuum state $|p,p^{*}>$ such that
\ber
\psi[r]|p,p^{*}>=\psi^{*}[r]|p,p^{*}>=0, r\geq {1\ov 2},\nmb
X[n]|p,p^{*}>=X^{*}[n]|p,p^{*}>=0, n\geq 1, \nmb
X[0]|p,p^{*}>=p|p,p^{*}>, \
X^{*}[0]|p,p^{*}>=p^{*}|p,p^{*}>.
\label{1.vac}
\enr
It is a primary state with respect
to the $N=2$ Virasoro algebra
\ber
G^{\pm}[r]|p,p^{*}>=0, r>0, \nmb
J[n]|p,p^{*}>=L[n]|p,p^{*}>=0, n>0, \nmb
J[0]|p,p^{*}>={j\ov \mu}|p,p^{*}>=0, \nmb
L[0]|p,p^{*}>={h(h+2)-j^{2}\ov 4\mu}|p,p^{*}>=0,
\label{1.hwv}
\enr
where $j=p^{*}-\mu p$, $h=p^{*}+\mu p$.

 The character $f_{p,p^{*}}(q,u)$ of the Fock module
$F_{p,p^{*}}$ is given by 
\ber
f_{p,p^{*}}(q,u)\equiv Tr_{F_{p,p^{*}}}(q^{L[0]-{c\ov 24}}u^{J[0]})=
q^{{h(h+2)-j^{2}\ov 4\mu}-{c\ov24}}u^{{j\ov \mu}}
{\Theta(q,u)\ov \eta(q)^{3}},
\label{1.fchc}
\enr
where the Jacoby theta-function
\ber
\Theta(q,u)=
q^{1\ov 8}\sum_{m\in Z}q^{{1\ov2}m^{2}}u^{-m}
\label{1.Jthet}
\enr
and the Dedekind eta-function
\ber
\eta(q)=q^{1\ov 24}\prod_{m=1}(1-q^{m})
\label{1.eta}
\enr
have been used.

 The $N=2$ Virasoro algebra has the following set of automorphisms
which is known as spectral flow ~\cite{SS}
\ber
G^{\pm}[r]\rightarrow G_{t}^{\pm}[r]\equiv G^{\pm}[r\pm t], \nmb
L[n]\rightarrow L_{t}[n]\equiv L[n]+t J[n]+t^{2}{c\ov 6}\dlt_{n,0},\
J[n]\rightarrow J_{t}[n]\equiv J[n]+t {c\ov 3}\dlt_{n,0},
\label{1.flow}
\enr
where $t\in Z$. 

 The spectral flow action on the free
fields can be easily described if we bosonize fermions $\psi^{*}, \psi$
\ber
\psi(z)=\exp(-y(z)), \ \psi^{*}(z)=\exp(+y(z)).
\label{1.fbos}
\enr
and introduce spectral flow vertex operator
\ber
U^{t}(z)=\exp(-t(y+{1\ov \mu}X^{*}-X)(z)).
\label{1.vflow}
\enr
It gives the action of spectral flow on the modes of the free-fields
\ber
\psi[r]\rightarrow \psi[r-t], \
\psi^{*}[r]\rightarrow \psi^{*}[r+t], \nmb
X^{*}[n]\rightarrow X^{*}[n]+t\dlt_{n,0}, \
X[n]\rightarrow X[n]-{t\ov \mu}\dlt_{n,0}.
\label{1.3}
\enr

 The action of the spectral flow on the vertex operator
$V_{(p,p^{*})}(z)$ is given by the normal ordered product of the
vertex $U^{t}(z)$ and $V_{p,p^{*}}(z)$. It follows from (\ref{1.3}) that
spectral flow generates twisted sectors.

\leftline{\bf 1.2. Irreducible $N=2$ super-Virasoro representations
and butterfly resolution.}

 The $N=2$ minimal models
are characterized by the condition that
$\mu$ is integer and $\mu\geq 2$. In NS sector the irreducible
highest-weight modules, constituting the (left-moving) space of states of
the minimal model, are unitary and labeled by two integers $h,j$,
where $h=0,...,\mu-2$ and $j=-h,-h+2,...,h$.
The highest-weight vector $|h,j>$ of the module satisfies the conditions
\ber
G^{\pm}[r]|h,j>=0, r>0, \nmb
J[n]|h,j>=L[n]|h,j>=0, n>0, \nmb
J[0]|h,j>={j\ov \mu}|h,j>, \nmb
L[0]|h,j>={h(h+2)-j^{2}\ov 4\mu}|h,j>.
\label{1.hw}
\enr

 The Fock modules are highly reducible representations of $N=2$ Virasoro algebra
and hence contain infinite number of singular vectors.
To describe the singular vectors structure we introduce following 
to ~\cite{FeS} the pair of fermionic screening currents
$S^{\pm}(z)$ and the screening charges $Q^{\pm}$
\ber
S^{+}(z)=\psi^{*}\exp(X^{*})(z), \
S^{-}(z)=\psi\exp(\mu X)(z), \nmb
Q^{\pm}=\oint dz S^{\pm}(z)
\label{1.chrg}
\enr
The screening charges commute with the generators of $N=2$ super-Virasoro algebra
(\ref{1.min}). 
But they do not act within each Fock module. Instead they relate to each other the
different Fock modules. The space where the screening charges are acting naturally 
is the direct sum of Fock modules
\ber
F_{\pi}=\oplus_{(p,p^{*})\in \pi}F_{p,p^{*}},
\label{1.fock}
\enr
where $\pi$ is the lattice of momentums:
\ber
\pi=\{(p,p^{*})|p={n\ov\mu},p^{*}=m, n,m\in Z\}.
\label{1.L}
\enr
Application of the screening charge to an arbitrary vector $|p,p^{*}>\in F_{\pi}$
gives the singular vector from another Fock module.

The screening charges are
nilpotent and mutually anti-commute
\ber
(Q^{+})^{2}=(Q^{-})^{2}=\{Q^{+},Q^{-}\}=0.
\label{1.BRST}
\enr
Due to this important properties (\ref{1.BRST}) one can combine the charges $Q^{\pm}$ into
BRST operator acting in $F_{\pi}$ and build a BRST complex of Fock modules 
$F_{p,p^{*}}\in F_{\pi}$. This complex 
which has been constructed in ~\cite{FeS} describes the structure
of N=2 Virasoro superalgebra singular vectors and corresponding submodules such that the
cohomology of the complex gives the irreducible module $M_{h,j}$.

 Let us consider first free-field construction
for the chiral module $M_{h,j=h}$. In this case the complex
(which is known due to Feigin and Semikhatov as butterfly resolution)
can be represented by the following diagram
\ber
\begin{array}{ccccccccccc}
&&\vdots &\vdots &&&&&&\\
&&\uparrow &\uparrow &&&&&&\\
\ldots &\leftarrow &F_{1,h+\mu} &\leftarrow
F_{0,h+\mu}&&&&&&\\
&&\uparrow &\uparrow &&&&&&\\
\ldots &\leftarrow &F_{1,h} &\leftarrow F_{0,h}&&&&&&\\
&&&&\nwarrow&&&&&\\
&&&&&F_{-1,h-\mu}&\leftarrow &F_{-2,h-\mu}&\leftarrow&\ldots\\
&&&&&\uparrow &&\uparrow&\\
&&&&&F_{-1,h-2\mu}&\leftarrow &F_{-2,h-2\mu}&\leftarrow&\ldots\\
&&&&&\uparrow &&\uparrow &&\\
&&&&&\vdots &&\vdots &&
\end{array} \nmb
\label{1.but}
\enr
We shall denote this resolution by $C_{h}$ and denote by $\Gm$
the set where the momentums of the Fock spaces of the resolution take
values.
The horizontal arrows in this diagram are given by the action of
$Q^{+}$ and vertical arrows are given by the action of $Q^{-}$.
The diagonal arrow at the middle of butterfly resolution
is given by the action of $Q^{+}Q^{-}$ (which equals $-Q^{-}Q^{+}$
due to (\ref{1.BRST})). Ghost number operator $g$ of the complex
is defined for an arbitrary vector $|v_{n,m}>\in
F_{n,m\mu+h}$ by
\ber
g|v_{n,m}>=(n+m)|v_{n,m}>, \ if \ n,m\geq 0, \nmb
g|v_{n,m}>=(n+m+1)|v_{n,m}>, \ if \ n,m< 0.
\label{1.grad}
\enr

The main statement of ~\cite{FeS} is that the complex (\ref{1.but}) is exact
except at the $F_{0,h}$ module, where the cohomology is given by
the chiral module $M_{h,j=h}$.

 The butterfly resolution allows to write the character
$\chi_{h}(q,u)\equiv Tr_{M_{h,h}}(q^{L[0]-{c\ov 24}}u^{J[0]})$ of the
module $M_{h,h}$ as the Euler characteristic of the complex:
\ber
\chi_{h}(q,u)=\chi^{(l)}_{h}(q,u)-\chi^{(r)}_{h}(q,u), \nmb
\chi^{(l)}_{h}(q,u)=\sum_{n,m\geq 0}(-1)^{n+m}f_{n,h+m\mu}(q,u),\nmb
\chi^{(r)}_{h}(q,u)=\sum_{n,m> 0}(-1)^{n+m}f_{-n,h-m\mu}(q,u),
\label{1.char}
\enr
where $\chi^{(l)}_{h}(q,u)$ and $\chi^{(r)}_{h}(q,u)$ are the characters
of the left and right wings of the resolution.

 To get the resolutions for other (anti-chiral and non-chiral) modules
one can use the observation ~\cite{FeS} that all irreducible modules can be obtained
from the chiral modules $M_{h,j=h}$, $h=0,...,\mu-2$ by the spectral flow
action $U^{-t}, t=h,h-1,...1$. Equivalently one can restrict the set of chiral
modules by the range $h=0,...,[{\mu\ov2}]-1$ and extend the spectral flow
action by
$t=\mu-1,...,1$ (when $\mu$ is even and $h=[{\mu\ov2}]-1$ the spectral flow
orbit becomes shorter: $t=[{\mu\ov2}]-1,...,1$)~\cite{FeSST}. Thus
the set of irreducible modules can be labeled also by the set
$\{(h,t)|h=0,...,[{\mu\ov2}]-1,\ t=\mu-1,...,0 \}$,
except the case when $\mu$ is even and the spectral flow orbit becomes
shorter.
It turns out that one can get
all the resolutions by the spectral flow action also.

 Due to this discussion it is more convenient to change the notation for irreducible
modules. In what follows we shall denote the irreducible modules as $M_{h,t}$,
indicating by $t$ spectral flow parameter.

 As well as the modules and resolutions one can get the characters by the
spectral flow action ~\cite{FeS}:
\ber
\chi_{h,t}(q,u)=q^{{c\ov6}t^{2}}u^{{c\ov3}t}
\chi_{h}(q,uq^{t}).
\label{1.tchi}
\enr
There are the following important automorphism properties of irreducible modules and characters
~\cite{FeS},~\cite{FeSST}.
\ber
M_{h,t}\equiv M_{\mu-h-2,t-h-1}, \
\chi_{h,t}(q,u)=\chi_{\mu-h-2,t-h-1}(q,u),
\label{1.reflect}
\enr
\ber
M_{h,t}\equiv M_{h,t+\mu}, \ \chi_{h,t+\mu}(q,u)=\chi_{h,t}(q,u),
\label{1.oddper}
\enr
where $\mu$ is odd and
\ber
M_{h,t}\equiv M_{h,t+\mu}, \ \chi_{h,t+\mu}(q,u)=\chi_{h,t}(q,u),\ h\neq [{\mu\ov2}]-1, \nmb
M_{h,t}\equiv M_{h,t+[{\mu\ov2}]}, \
\chi_{h,t+[{\mu\ov2}]}(q,u)=\chi_{h,t}(q,u),\ h=[{\mu\ov2}]-1,
\label{1.evper}
\enr
where $\mu$ is even.

  Note that the butterfly resolution
is not periodic under the spectral flow as opposed to the
characters. It is also not invariant with respect to the automorphism (\ref{1.reflect}).
Instead, the periodicity and invariance are recovered on the level of
cohomology. Thus, $U^{\pm\mu}$ spectral flow and automorphism (\ref{1.reflect}) are the
quasi-isomorphisms of complexes.

 The modules, resolutions and characters in R sector are
generated from the modules, resolutions and characters in NS sector by the
spectral flow operator $U^{-{1\ov2}}$.

\vskip 10pt
\centerline{\bf 2. Free-field realization of Gepner model.}
\vskip 10pt

\leftline {\bf 2.1.Free-field realization of the product of
minimal models.}

 It is easy to generalize the free-field representation of the Sect.1. to the case of
tensor product of $r$ $N=2$ minimal models which can be characterized
by $r$ dimensional vector $\bmu=(\mu_{1},...,\mu_{r})$,
where $\mu_{i}\geq 2$ and integer.

 Let $E$ be a real $r$ dimensional vector space and let $E^{*}$
be the dual space to $E$. Let us denote by $<,>$
the natural scalar product in the direct sum $E \oplus E^{*}$.
In the subspaces $E$ and $E^{*}$ we fix the sets of basic vectors
$\Re$ and $\Re^{*}$
\ber
\Re=\{{\bf s}_{i}, i=1,...,r\},
\nmb
\Re^{*}=\{\mu_{i}{\bf s}^{*}_{i}, i=1,...,r\},
\nmb
<{\bf s}_{i},{\bf s}^{*}_{j}>=\dlt_{i,j}.
\label{2.bases}
\enr

 According to the $\Re$ and $\Re^{*}$
we introduce (in the left-moving sector)
the free bosonic fields $X_{i}(z), X^{*}_{i}(z)$ and free
fermionic fields $\psi_{i}(z), \psi^{*}_{i}(z)$, $i=1,...,r$
so that its singular OPE's are given by (\ref{1.ope}) as well as
the following fermionic screening currents and their charges
\ber
S^{+}_{i}(z)={\bf s}_{i}\psi^{*}\exp({\bf s}_{i}X^{*})(z), \nmb
S^{-}_{i}(z)={\bf s}^{*}_{i}\psi\exp(\mu_{i}{\bf s}^{*}_{i}X)(z), \nmb
Q^{\pm}_{i}=\oint dz S^{\pm}_{i}(z).
\label{2.screen}
\enr
For each $i=1,...,r$ one can define by the formulas (\ref{1.min})
N=2 $c_{i}=3(1-{2\ov \mu_{i}})$
Virasoro superalgebra
\ber
G^{+}_{i}={\bf s}_{i}\psi^{*} {\bf s}^{*}_{i}\d X
-{1\ov \mu_{i}}{\bf s}_{i}\d \psi^{*}, \
G^{-}_{i}={\bf s}^{*}_{i}\psi {\bf s}_{i}\d X^{*}
-{\bf s}^{*}_{i}\d \psi, \nmb
J_{i}=
({\bf s}_{i}\psi^{*}{\bf s}^{*}_{i}\psi+
{1\ov \mu_{i}}{\bf s}_{i}\d X^{*}-{\bf s}^{*}_{i}\d X), \nmb
T_{i}(z)=
{1\ov 2}({\bf s}_{i}\d \psi^{*}{\bf s}^{*}_{i}\psi-
{\bf s}_{i}\psi^{*}{\bf s}^{*}_{i}\d \psi)+
{\bf s}_{i}\d X^{*}{\bf s}^{*}_{i}\d X-  \nmb
{1\ov 2}({\bf s}^{*}_{i}\d^{2} X+{1\ov \mu_{i}}{\bf s}_{i}\d^{2} X^{*})
\label{2.min}
\enr
as well as the vertex operators
\ber
V_{(p_{i},p^{*}_{i})}=
\exp(p^{*}_{i}{\bf s}^{*}_{i}X+p_{i}{\bf s}_{i}X^{*})),
\label{2.V}
\enr
which are the conformal fields whose conformal dimensions and charges
are labeled by integers $h_{i}=p^{*}_{i}+\mu_{i} p_{i}$, 
$j_{i}=p^{*}_{i}-\mu_{i} p_{i}$.

 The vertex operators are naturally associated to the
lattice $\Pi=P\oplus P^{*}\in E\oplus E^{*}$,
where $P\in E, P^{*}\in E^{*}$ such that
$P$ is generated by ${1\ov \mu_{i}}{\bf s}_{i}$ and
$P^{*}$ is generated by the basis
${\bf s}^{*}_{i}$, $i=1,...,r$.
For an arbitrary vector $(\bp,\bp^{*})\in\Pi$,
we introduce in NS sector Fock vacuum state $|\bp,\bp^{*}>$ by the
formulas similar to (\ref{1.vac}) and denote by $F_{\bp,\bp^{*}}$ the Fock module
generated from $|\bp,\bp^{*}>$ by the creation operators of the fields
$X_{i}(z), X^{*}_{i}(z)$, $\psi_{i}(z), \psi^{*}_{i}(z)$.

 Let $F_{\Pi}$ be the direct sum of Fock modules associated to the lattice $\Pi$.
As an obvious generalization of the results
from Sec.1. we form for each vector $\bh=\sum_{i}h_{i}{\bf s}^{*}_{i}\in P^{*}$,
where $h_{i}=0,1,...,\mu_{i}-2$ butterfly resolution
$C^{\star}_{\bh}$ as the product $\otimes_{i=1}^{r}C^{\star}_{h_{i}}$ of butterfly resolutions
of minimal models.
The corresponding ghost number operator $g$ is given by the sum of
ghost number operators of each of the resolutions. The
differential $\d$ acting on ghost number
$N$ subspace of the resolution
is given by the sum of differentials of each of the complexes $C^{\star}_{h_{i}}$.
It is obvious that the complex $C^{\star}_{\bh}$ is exact
except at the $F_{0,\bh}$ module, where the cohomology is given by the product
$M_{\bh,0}=\otimes_{i=1}^{r}M_{h_{i},0}$ of the chiral modules
of each minimal model. Hence one can represent the character
\ber
\chi_{\bh,0}(q,u)\equiv Tr_{M_{\bh,0}}(q^{L[0]-{c\ov24}})u^{J[0]})
\label{2.char}
\enr
of $M_{\bh,0}$
as the product of characters $\chi_{\bh,0}(q,u)=\prod_{i}\chi_{h_{i},0}(q,u)$.

 According to the discussion at the end of Sec.1. we obtain the
resolution and character for the product
of arbitrary irreducible modules of minimal models acting on $C^{\star}_{\bh}$ by the
spectral flow operators $U^{-\bt}=\prod_{i}U_{i}^{-t_{i}}$ of the minimal
models. Hence one can label the resolutions, modules and characters by the pairs of 
vectors $(\bh,\bt)$, from the
set $\tld{\Dl}=\{(\bh,\bt)|h_{i}=0,...,[{\mu_{i}\ov2}]-1,\
t_{i}=0,...,\mu_{i}-1,\ i=1,...,r\}$. On the equal footing one can use the set
$\tld{\Dl}'=\{(\bh,\bt)|h_{i}=0,...,\mu_{i}-2,\
t_{i}=0,...,h_{i},\ i=1,...,r\}$.

 It is also clear that R sector resolutions, modules and characters are generated
from NS sector by the spectral flow $U^{-\bv/2}=\prod_{i=1}^{r}U_{i}^{-1/2}$,
where $\bv=(1,...,1)$.

 The same free-field realization can be used in the
right-moving sector. Thus the sets of screening
vectors  $\bar{\Re}$ and $\bar{\Re}^{*}$ have to be fixed in the
right-moving sector. It can be done in many ways, the only
restriction is that the corresponding cohomology group has to be
isomorphic to the space of states of the product of minimal models in the
right-moving sector. Therefore $\bar{\Re}$ and $\bar{\Re}^{*}$
are determined modulo $O(r,r)$ transformations which left unchanged
the matrix of scalar products $<{\bf s}_{i},{\bf s}^{*}_{j}>$.
In what follows we
put
\ber
\bar{\Re}=\Re, \ \bar{\Re}^{*}=\Re^{*}.
\label{2.bbases}
\enr
Hence, one can use the
same complex to describe the irreducible modules in the
right-moving sector.

\leftline {\bf 2.2.Free-field realization and Calabi-Yau extension.}

 It is well known that product of minimal models can not be
applied straightforward to describe the string theory on
$2D$-dimensional CY manifold. First, one has to demand
that $\sum_{i}c_{i}=3D$. Second, the so called simple current
orbifold $CY_{\bmu}$ ~\cite{EOTY}, ~\cite{SCORB}, ~\cite{FSW} of the product 
of minimal models has to be constructed. The currents
of $N=2$ Virasoro superalgebra of this model are given by the sum of
currents of each minimal model
\ber
G^{\pm}(z)=\sum_{i}G^{\pm}_{i}, \nmb
J(z)=\sum_{i}J_{i}, \
T(z)=\sum_{i}T_{i}.
\label{2.Vird}
\enr

 The left-moving (as well as the right-moving) sector of the $CY_{\bmu}$ is given by
projection of the space of states on the
subspace of integer $J[0]$-charges and organizing the projected
space into the orbits $[\bh,\bt]$ under the spectral flow operator
$U^{\bv}=\prod_{i=1}^{r}U_{i}$ ~\cite{FSW}.

 The partition function in NS sector of $CY_{\bmu}$ sigma model
is diagonal modular invariant of the spectral flow orbits characters restricted to the
subset of integer $J[0]$ charges. From $N=2$ Virasoro superalgebra representations there is no
difference what of the sets $\tld{\Dl}$ or $\tld{\Dl}'$ we use to parameterize the orbit characters (though their
free-field realizations are different). In what follows
we combine these to sets into the
extended set $\Dl=\{(\bh,\bt)|h_{i}=0,...,\mu_{i}-2,\
t_{i}=0,...,\mu_{i}-1,\ i=1,...,r\}$
and take into account this extension by corresponding multiplier 
("field identification") ~\cite{Gep}.

 The orbit characters (with the restriction on integer charges
subspace) can be written in explicit form so that the structure of simple
current extension becomes clear ~\cite{EOTY},~\cite{FSW}:
\ber
ch_{\bh,\bt}(q,u)={1\ov\kp^{2}}\sum_{n,m=0}^{\kp-1}
Tr_{M_{\bh,\bt}}(U^{n\bv}q^{(L[0]-{c\ov24})}u^{J[0]}\exp{(\im2\pi
mJ[0])}U^{-n\bv})=
\nmb
{1\ov\kp^{2}}\sum_{n,m=0}^{\kp-1}\chi_{\bh,\bt+n\bv}(\tau,\ups+m),
\label{2.orbchi}
\enr
where $q=\exp{(\im2\pi\tau)}$, $u=\exp{(\im2\pi\ups)}$ and
$\kp=lcm\{\mu_{i}\}$.
The partition function of $CY_{\bmu}$ model is given by
\ber
Z_{CY}(q,\bar{q})={1\ov 2^{r}}\sum_{[\bh,\bt]\in \Dl_{CY}}\kp|ch_{[\bh,\bt]}(q)|^{2},
\label{2.ZCY}
\enr
where $\Dl_{CY}$ denotes the subset of $\Dl$ restricted to the space of integer $J[0]$ charges.
$[\bh,\bt]$ denotes the spectral flow orbit of the point $(\bh,\bt)$.
Factor ${1\ov 2^{r}}$ corresponds to the extended set $\Dl$ of irreducible modules and
$\kp$ is the length of the orbit $[\bh,\bt]$. In general case the
orbits with different lengths could appear but we will not consider these
cases to escape the problem of fixed point resolution ~\cite{SCORB}, ~\cite{FSW},
~\cite{BRS}.

\leftline {\bf 2.3.Free-field realization of Gepner
models.}

 The Gepner models ~\cite{Gep} of CY superstring
compactification are given by (generalized) GSO projection ~\cite{Gep}, ~\cite{EOTY}
which is carrying out on the product of the space of states
of $CY_{\bmu}$ model and space of states of external fermions and
bosons describing space-time degrees of freedom of the string.
In the framework of simple current extension formalism the Gepner's construction
has been farther developed in ~\cite{FSW},
~\cite{FSS}, ~\cite{SCORB}.

 Let us introduce so called supersymmetrized (Green-Schwartz)
characters ~\cite{Gep},~\cite{EOTY}
\ber
Ch_{[\bh,\bt]}(q,u)={1\ov 4\kp^{2}}\sum_{n,m=0}^{2\kp-1}
Tr_{(M_{\bh,\bt}\otimes\Phi)}(U_{tot}^{m\ov2}\exp{(\im\pi nJ_{tot}[0])}
q^{(L_{tot}-{c_{tot}\ov24})}u^{J_{tot}[0]}U_{tot}^{-m\ov2}),
\label{2.gschar}
\enr
where the trace is calculated in the product of
$M_{\bh,\bt}$ and Fock module $\Phi$
generated by the external (space-time) fermions and bosons in NS sector,
$J_{tot}[0]$ and $L_{tot}[0]$ are zero modes of the total $U(1)$ current
and stress-energy tensor which includes the contributions from space-time
degrees of freedom, $c_{tot}=c+{3\ov2}(8-2D)=12$ is a total central charge and
$U_{tot}$ is a total spectral flow operator acting
in the product $M_{\bh,\bt}\otimes\Phi$.

 The modular invariant Gepner model partition function is given by
~\cite{Gep},~\cite{EOTY},~\cite{FSW}
\ber
Z_{Gep}(q,\bar{q})={1\ov2^{r}}(Im\tau)^{-(4-D/2)}
\sum_{[\bh,\bt]\in \Dl_{CY}}\kp|Ch_{[\bh,\bt]}(q)|^{2}.
\label{2.ZG}
\enr

\vskip 10pt
\centerline{\bf 3. The Ishibashi states in Fock modules.}
\vskip 10pt

 The boundary states we are going to construct can be considered
as a bilinear forms on the space of states of the model. Thus, it
will be implied in what follows that the right-moving sector of
the model is realized by the free-fields
$\bar{X}_{i}(\bar{z}), \bar{X}^{*}_{i}(\bar{z}), \bar{\psi}_{i}(\bar{z}),
\bar{\psi}^{*}_{i}(\bar{z})$, $i=1,...,r$ and the right-moving
$N=2$ super-Virasoro algebra is given by the formulas similar to
(\ref{1.min})

 There are two types of boundary states preserving $N=2$
super-Virasoro algebra ~\cite{OOY}, usually called $B$-type
\ber
(L[n]-\bar{L}[-n])|B>>=(J[n]+\bar{J}[-n])|B>>=0, \nmb
(G^{+}[r]+\imath \et \bar{G}^{+}[-r])|B>>=
(G^{-}[r]+\imath \et \bar{G}^{-}[-r])|B>>=0
\label{3.BD}
\enr
and $A$-type states
\ber
(L[n]-\bar{L}[-n])|A>>=(J[n]-\bar{J}[-n])|A>>=0, \nmb
(G^{+}[r]+\im \et \bar{G}^{-}[-r])|A>>=
(G^{-}[r]+\im \et \bar{G}^{+}[-r])|A>>=0
\label{3.AD}
\enr
where $\et=\pm 1$.

 In the tensor product
of the left-moving Fock module $F_{\bp,\bp^{*}}$ and right-moving Fock
module $\bar{F}_{\bar{\bp},\bar{\bp}^{*}}$ we construct the most
simple states fulfilling the solutions (\ref{3.BD}) and
(\ref{3.AD}). We shall call these states as Fock space Ishibashi~\cite{Ish}
states.
 
\leftline{\bf 3.1. B-type Ishibashi states in Fock module.}

 Let us consider in NS sector the following ansatz for fermions
\ber
(\psi^{*}_{i}[r]-
\im \et \Om_{ij}\bar{\psi}^{*}_{j}[-r])|\bp,\bp^{*},\bar{\bp},\bar{\bp}^{*},\et,B>>=0,
\nmb
(\psi_{i}[r]-
\im \et \Om^{*}_{ij}\bar{\psi}_{j}[-r])|\bp,\bp^{*},\bar{\bp},\bar{\bp}^{*},\et,B>>=0
\label{3.Bantz}
\enr
where $\Om_{ij},\Om^{*}_{ij}$ are the arbitrary nondegenerate matrices.
Substituting these relations
into (\ref{3.BD}) and using (\ref{2.min}), (\ref{2.Vird}) we find
\ber
\Om_{ik}\Om^{*}_{in}=\dlt_{kn},\nmb
\Om_{ij}d_{i}=d_{j}, \
\Om^{*}_{ij}d^{*}_{i}=d^{*}_{j}\nmb
\bar{p}_{k}=-\Om_{jk}p_{j}-d_{k}, \
\bar{p}^{*}_{k}=-\Om^{*}_{jk}p^{*}_{j}-d^{*}_{k}, \nmb
(\Om_{jk}X_{j}[n]+\bar{X}_{k}[-n]+d_{k}\dlt_{n,0})
|\bp,\bp^{*},\bar{\bp},\bar{\bp}^{*},\et,B>>=0, \nmb
(\Om^{*}_{jk}X^{*}_{j}[n]+\bar{X}^{*}_{k}[-n]+d^{*}_{k}\dlt_{n,0})
|\bp,\bp^{*},\bar{\bp},\bar{\bp}^{*},\et,B>>=0,
\label{3.BX}
\enr
where $d_{k}={1\ov\mu_{k}}$, $d^{*}_{k}=1$ and we combine these coefficients into
the $r$-dimensional vectors
$\bd=(d_{1},...,d_{r})$, $\bd^{*}=(d^{*}_{1},...,d^{*}_{r})$.

 It is helpful to rewrite the boundary conditions in
toric coordinates on the target space:
\ber
\theta_{i}[n]={\im\ov\sqrt{2\mu_{i}}}(X^{*}_{i}[n]-\mu_{i} X_{i}[n]), \
R_{i}[n]={1\ov\sqrt{2\mu_{i}}}(X^{*}_{i}[n]+\mu_{i} X_{i}[n]), \nmb
\gm_{i}[s]={\im\ov\sqrt{2\mu_{i}}}(\psi^{*}_{i}[s]-\mu_{i}\psi_{i}[s]),\
\sgm_{i}[s]={1\ov\sqrt{2\mu_{i}}}(\psi^{*}_{i}[s]+\mu_{i}\psi_{i}[s]).
\label{3.tor}
\enr

Then (\ref{3.Bantz}) and (\ref{3.BX}) take the form
\ber
(\sgm_{i}[s]-{\im\et\ov2}(\sqrt{{\mu_{j}\ov\mu_{i}}}\Om_{ij}+
\sqrt{{\mu_{i}\ov\mu_{j}}}\Om^{*}_{ij})\bar{\sgm}_{j}[-s]-
{\et\ov2}(\sqrt{{\mu_{j}\ov\mu_{i}}}\Om_{ij}-
\sqrt{{\mu_{i}\ov\mu_{j}}}\Om^{*}_{ij})\bar{\gm}_{j}[-s])|B>>=0, \nmb
(\gm_{i}[s]+{\et\ov2}(\sqrt{{\mu_{j}\ov\mu_{i}}}\Om_{ij}-
\sqrt{{\mu_{i}\ov\mu_{j}}}\Om^{*}_{ij})\bar{\sgm}_{j}[-s]-
{\im\et\ov2}(\sqrt{{\mu_{j}\ov\mu_{i}}}\Om_{ij}+
\sqrt{{\mu_{i}\ov\mu_{j}}}\Om^{*}_{ij})\bar{\gm}_{j}[-s])|B>>=0,
\nmb
(\bar{R}_{j}[-n]+{1\ov2}(\sqrt{{\mu_{i}\ov\mu_{j}}}\Om^{*}_{ij}+
\sqrt{{\mu_{j}\ov\mu_{i}}}\Om_{ij})R_{i}[n]-
{\im\ov2}(\sqrt{{\mu_{i}\ov\mu_{j}}}\Om^{*}_{ij}-
\sqrt{{\mu_{j}\ov\mu_{i}}}\Om_{ij})\theta_{i}[n]+
\sqrt{{2\ov\mu_{j}}}\dlt_{n,0})|B>>=0,
\nmb
(\bar{\theta}_{j}[-n]+{\im\ov2}(\sqrt{{\mu_{i}\ov\mu_{j}}}\Om^{*}_{ij}-
\sqrt{{\mu_{j}\ov\mu_{i}}}\Om_{ij})R_{i}[n]+
{1\ov2}(\sqrt{{\mu_{i}\ov\mu_{j}}}\Om^{*}_{ij}+\sqrt{{\mu_{j}\ov\mu_{i}}}\Om_{ij})
\theta_{i}[n])|B>>=0.
\label{3.Btor}
\enr
Because of
the toric coordinates $(\theta_{i},R_{i})$ are real we have to put
the reality constraint 
\ber
\Om^{*}_{ij}={\mu_{j}\ov\mu_{i}}\bar{\Om}_{ij}.
\label{3.realB}
\enr

 The linear $B$-type Ishibashi state in NS sector is given by the
standard expression ~\cite{CLNY},~\cite{PC}
\ber
|\bp,\bp^{*},\Om,\et,B>>=
\prod_{n=1}\exp(-{1\ov n}(X^{*}_{i}[-n]\Om^{*}_{ik}\bar{X}_{k}[-n]+
X_{i}[-n]\Om_{ik}\bar{X}^{*}_{k}[-n])) \nmb
\prod_{r=1/2}\exp(\im\et(\psi^{*}_{i}[-r]\Om^{*}_{ik}\bar{\psi}_{k}[-r]+
\psi_{i}[-r]\Om_{ik}\bar{\psi}^{*}_{k}[-r]))
|\bp,\bp^{*},-\Om^{T}\bp-\bd,-(\Om^{*})^{T}\bp^{*}-\bd^{*}>.
\label{3.BI}
\enr

\leftline{\bf 3.2. A-type Ishibashi states in Fock module.}

 The $A$-type Ishibashi states in Fock module can be found analogously. The
linear ansatz for fermions has the form
\ber
(\psi^{*}_{i}[r]-
\im \et \Ups_{ij}\bar{\psi}_{j}[-r])|\bp,\bp^{*},\bar{\bp},\bar{\bp}^{*},\et,A>>=0,
\nmb
(\psi_{i}[r]-
\im \et \Ups^{*}_{ij}\bar{\psi}^{*}_{j}[-r])|\bp,\bp^{*},\bar{\bp},\bar{\bp}^{*},\et,A>>=0
\label{3.Aantz}
\enr
where $\Ups_{ij},\Ups^{*}_{ij}$ are the arbitrary nondegenerate matrices.
Substituting these relations
into (\ref{3.AD}) and using (\ref{1.min}) we find
\ber
\Ups_{ik}\Ups^{*}_{in}=\dlt_{kn},\nmb
\Ups_{ij}d_{i}=d_{j}^{*}, \
\Ups^{*}_{ij}d_{j}^{*}=d_{i},
\nmb
\bar{p}_{k}=-\Ups^{*}_{jk}p^{*}_{j}-d_{k}, \
\bar{p}^{*}_{k}=-\Ups_{jk}p_{j}-d^{*}_{k}, \nmb
(\Ups_{jk}X_{j}[n]+\bar{X}^{*}_{k}[-n]+d_{k}^{*}\dlt_{n,0})
|\bp,\bp^{*},\bar{\bp},\bar{\bp}^{*},\et,A>>=0, \nmb
(\Ups^{*}_{jk}X^{*}_{j}[n]+\bar{X}_{k}[-n]+d_{k}\dlt_{n,0})
|\bp,\bp^{*},\bar{\bp},\bar{\bp}^{*},\et,A>>=0.
\label{3.AX}
\enr

 In the toric coordinates (\ref{3.tor}) the conditions take the
form
\ber
(\sgm_{i}[s]-{\im\et\ov2}({\Ups_{ij}\ov\sqrt{\mu_{j}\mu_{i}}}+
\sqrt{\mu_{j}\mu_{i}}\Ups^{*}_{ij})\bar{\sgm}_{j}[-s]+
{\et\ov2}({\Ups_{ij}\ov\sqrt{\mu_{i}\mu_{j}}}-
\sqrt{\mu_{i}\mu_{j}}\Ups^{*}_{ij})\bar{\gm}_{j}[-s])|A>>=0, \nmb
(\gm_{i}[s]+{\et\ov2}({\Ups_{ij}\ov\sqrt{\mu_{i}\mu_{j}}}-
\sqrt{\mu_{i}\mu_{j}}\Ups^{*}_{ij})\bar{\sgm}_{j}[-s]+
{\im\et\ov2}({\Ups_{ij}\ov\sqrt{\mu_{j}\mu_{i}}}+
\sqrt{\mu_{j}\mu_{i}}\Ups^{*}_{ij})\bar{\gm}_{j}[-s])|A>>=0, \nmb
(\bar{R}_{j}[-n]+{1\ov2}({\Ups_{ij}\ov\sqrt{\mu_{j}\mu_{i}}}+
\sqrt{\mu_{j}\mu_{i}}\Ups^{*}_{ij})R_{i}[n]+
{\im\ov2}({\Ups_{ij}\ov\sqrt{\mu_{i}\mu_{j}}}-
\sqrt{\mu_{i}\mu_{j}}\Ups^{*}_{ij})\theta_{i}[n]+
2d^{*}_{j}\dlt_{n,0})|A>>=0, \nmb
(\bar{\theta}_{j}[-n]+{\im\ov2}
({\Ups_{ij}\ov\sqrt{\mu_{i}\mu_{j}}}-
\sqrt{\mu_{i}\mu_{j}}\Ups^{*}_{ij})R_{i}[n]-
{1\ov2}({\Ups_{ij}\ov\sqrt{\mu_{j}\mu_{i}}}+
\sqrt{\mu_{i}\mu_{j}}\Ups^{*}_{ij})\theta_{i}[n])|A>>=0,
\label{3.Ator}
\enr
The reality constraint takes the form
\ber
\Ups^{*}_{ij}={1\ov\mu_{i}\mu_{j}}\bar{\Ups}_{ij}.
\label{3.realA}
\enr

 The linear $A$-type Ishibashi state in NS sector is given similar
to $B$-type
\ber
|\bp,\bp^{*},\Ups,\et,A>>=
\prod_{n=1}\exp(-{1\ov
n}(X_{i}[-n]\Ups_{ik}\bar{X}_{k}[-n]+
X^{*}_{i}[-n]\Ups^{*}_{ik}\bar{X}^{*}_{k}[-n])) \nmb
\prod_{r=1/2}\exp(\im\et(\psi_{i}[-r]\Ups_{ik}\bar{\psi}_{k}[-r]+
\psi^{*}_{i}[-r]\Ups^{*}_{ik}\bar{\psi}^{*}_{k}[-r]))
|\bp,\bp^{*},-(\Ups^{*})^{T}\bp^{*}-\bd,-\Ups^{T}\bp-\bd^{*}>.
\label{3.AI}
\enr

\vskip 10pt
\centerline {\bf 4. Permutation Ishibashi states in the product of minimal models.}
\vskip 10pt
\leftline{\bf 4.1. B-type permutation Ishibashi states.}

 Free-field realizations of the irreducible modules described in Sect. 1,2
and the constructions (\ref{3.BI}), (\ref{3.AI})
allows to suggest that Ishibashi states in the product of minimal models can also 
be represented by the free-fields. Let us consider the following superposition of 
$B$-type Fock modules
Ishibashi states (\ref{3.BI})
\ber
|I_{\bh},\Om,\et,B>>=
\dlt(\Om\bh-\bh)\sum_{(\bp,\bp^{*})\in\Gm_{\bh}}c_{\bp,\bp^{*}}|\bp,\bp^{*},\Om,\et,B>>,
\label{4.Isuper}
\enr
where the coefficients $c_{\bp,\bp^{*}}$ are arbitrary and the summation is
performed over the momentums of the butterfly resolution
$C^{\star}_{\bh}$.  Since the partition function is diagonal the delta-function
$\dlt(\Om\bh-\bh)$ has been inserted. It is clear that this state satisfies the 
relations (\ref{3.BD}). 

 Before GSO projection, the closed string states of the model
which can interact with the Ishibashi state
(\ref{4.Isuper}) come from the product of left-moving and right-moving
Fock modules
$F_{\bp,\bp^{*}}\otimes
\bar{F}_{-\Om^{T}\bp-\bd,-(\Om^{*})^{T}\bp^{*}-\bd^{*}}$, 
where $(\bp,\bp^{*})\in\Gm_{\bh}$. The left-moving modules of the superposition
(\ref{4.Isuper}) constitute the
butterfly resolution $C_{\bh}$ whose cohomology is given by
the module $M_{\bh}$. What about the Fock modules from
the right-moving sector? To have nontrivial interaction with the states
from the model the right-moving Fock modules
have to from the product of resolutions of minimal models
(\ref{1.but}) also. But this contradicts to the relations between
left-moving and right-moving momentums from (\ref{3.BX}). 
This
contradiction may be resolved if we allow that right-moving Fock
modules form the product of resolutions each of which is dual
to (\ref{1.but}). The dual resolution $\tld{C}_{h}$ to the minimal model
resolution (\ref{1.but})
is given by the following diagram
\ber
\begin{array}{ccccccccccc}
&&\vdots &\vdots &&&&&&\\
&&\downarrow &\downarrow &&&&&&\\
\ldots &\rightarrow &\bar{F}_{-1-{1\ov\mu},-1-h-\mu} &\rightarrow
\bar{F}_{-{1\ov\mu},-1-h-\mu}&&&&&&\\
&&\downarrow &\downarrow &&&&&&\\
\ldots &\rightarrow &\bar{F}_{-1-{1\ov\mu},-1-h} &\rightarrow
\bar{F}_{-{1\ov\mu},-1-h}&&&&&&\\
&&&&\searrow&&&&&\\
&&&&&\bar{F}_{1-{1\ov\mu},-1-h+\mu}&\rightarrow &\bar{F}_{2-{1\ov\mu},-1-h+\mu}
&\rightarrow&\ldots\\
&&&&&\downarrow &&\downarrow&\\
&&&&&\bar{F}_{1-{1\ov\mu},-1-h+2\mu}&\rightarrow &\bar{F}_{2-{1\ov\mu},-1-h+2\mu}
&\rightarrow&\ldots\\
&&&&&\downarrow &&\downarrow &&\\
&&&&&\vdots &&\vdots &&
\end{array} \nmb
\label{3.dualbut}
\enr
(here, $h$ is
an integer number taking values from 0 to $\mu-2$)
The arrows on this diagram are given by the same operators as on
the diagram (\ref{1.but}).

 Hence the right-moving Fock modules have to form dual resolution
$\tld{C}_{\bh}=\otimes_{i=1}^{r}\tld{C}_{h_{i}}$ and matrices
$\Om^{T}$, $(\Om^{*})^{T}$ have to map the set of left-moving 
momentums $\Gm_{\bh}$ on
the set of momentums $\bar{\Gm}_{\bh}$ which has to be isomorphic to
$\Gm_{\bh}$. 
Therefore we conclude that $\Om^{T}$
has to be an element of the direct product of permutation groups
$\aleph _{r_{i}}$ of $r_{i}$-elements
\ber
\Om\in \aleph _{r_{1}...r_{N}}=\aleph _{r_{1}}\otimes \aleph _{r_{2}}...\otimes 
\aleph _{r_{N}},
\label{4.Bmatr}
\enr
which are determined by the sets $r_{1},...,r_{N}$ of coinciding elements
in the vector $\bmu$. In other words, it is implied here that we have
$\mu_{1}=...=\mu_{r_{1}}$, $\mu_{r_{1}+1}=...=\mu_{r_{1}+r_{2}}$,....
In view of (\ref{3.realB}) we have also
\ber
\Om^{*}_{ij}=\Om_{ij}.
\label{4.Bmatr1}
\enr
Thus the relations (\ref{3.Btor}) take the form
\ber
(\sgm_{i}[s]-\im\et \Om_{ij}\bar{\sgm}_{j}[-s])|B>>=0, \nmb
(\gm_{i}[s]-\im\et \Om_{ij}\bar{\gm}_{j}[-s])|B>>=0,
\nmb
(\bar{R}_{j}[-n]+\Om_{ij}R_{i}[n]+
\sqrt{{2\ov\mu_{j}}}\dlt_{n,0})|B>>=0,
\nmb
(\bar{\theta}_{j}[-n]+\Om_{ij}\theta_{i}[n])|B>>=0.
\label{4.Btorr}
\enr
Hence, $i$-th minimal model in the right-moving sector
interacts to $\Om^{-1}(i)$-th minimal model from the left-moving
sector.

 Having the matrix $\Om$ fixed by (\ref{4.Bmatr}) one can define
the coefficients $c_{\bp,\bp^{*}}$ from the $BRST$ invariance
condition. It is a straightforward generalization of the condition
found in ~\cite{SPmm} for $N=2$ minimal models. To formulate this
condition one has to describe by the free fields the total space
of states of the model.

 To do that we form first the product of complexes $C_{\bh}\otimes
\tld{C}_{\bh}$ to build the complex
\ber
\ldots\rightarrow {\bf C}_{\bh}^{-2}\rightarrow {\bf C}_{\bh}^{-1}
\rightarrow {\bf C}_{\bh}^{0}\rightarrow {\bf C}_{\bh}^{+1}\rightarrow\ldots,
\label{4.complex}
\enr
which is graded by the sum of the ghost numbers
$g+\bar{g}$ and for an arbitrary ghost number $I$
the space ${\bf C}_{\bh}^{I}$ is given by the sum of products of the
Fock modules from the resolution $C_{\bh}$ and $\tld{C}_{\bh}$
such that $g+\bar{g}=I$. The differential $\dlt$ of the complex
is defined by the differentials $\d$ and
$\bar{\d}$ of the complexes $C^{\star}_{\bh}$ and $\tld{C}^{\star}_{\bh}$
\ber
\dlt|v_{g}\otimes\bar{v}_{\bar{g}}>=|\d v_{g}\otimes\bar{v}_{\bar{g}}>+
(-1)^{g}|v_{g}\otimes\bar{\d}\bar{v}_{\bar{g}}>,
\label{3.Diff}
\enr
where $|v_{g}>$ is an arbitrary vector from the complex $C_{h}$
with ghost number $g$, while $|\bar{v}_{\bar{g}}>$ is an
arbitrary vector from the complex $\tld{C}^{\star}_{h}$ with the ghost
number $\bar{g}$ and $g+\bar{g}=I$.
The cohomology
of the complex (\ref{4.complex}) is nonzero only at grading 0
and is given by the product of irreducible modules
$M_{\bh}\otimes\bar{M}_{\bh,\bt=2\bh}$, where $\bar{M}_{\bh,\bt=2\bh}$
is the product of anti-chiral modules of minimal models.

 The Ishibashi state we are looking for
can be considered as a linear functional on the Hilbert space of the product of
models, then it has to be an element
of the homology group. Therefore, the $BRST$ invariance
condition for the state can be formulated as follows.

 Let us define the action of the differential $\dlt$ on the state
$|I_{\bh},\Om,\et,B>>$ by the formula
\ber
<<\dlt^{*}(I_{\bh},\Om,\et,B)|v_{g}\otimes \bar{v}_{\bar{g}}>\equiv
<<I_{\bh},\Om,\et,B|\dlt_{g+\bar{g}}|v_{g}\otimes \bar{v}_{\bar{g}}>,
\label{4.D*}
\enr
where $v_{g}\otimes \bar{v}_{\bar{g}}$ is an arbitrary element
from ${\bf C}_{\bh}^{g+\bar{g}}$. Then, $BRST$ invariance condition
means that
\ber
\dlt^{*}|I_{\bh},\Om,\et,B>>=0.
\label{4.Dcycl}
\enr

 As a straightforward generalization of Theorem 2 from ~\cite{SPmm}
we find that superposition (\ref{4.Isuper}) satisfies
$BRST$ invariance condition (\ref{4.Dcycl}) if the coefficients
$c_{\bp,\bp^{*}}$ take values $\pm 1$ according to  the expression
\ber
c_{\bp,\bp^{*}}=\sqrt{2}\cos((2g_{\bp,\bp^{*}}+1){\pi\ov4})c_{0,\bh},
\label{3.c}
\enr
where $g_{\bp,\bp^{*}}$ is the ghost number.

Thus the superposition (\ref{4.Isuper}) respects the singular vector structure
of the product of minimal N=2 Virasoro algebra representations and gives explicit
construction of permutation Ishibashi states.
Note also that $BRST$ condition doesn't fix the phase of the overall
coefficient $c_{0,\bh}$.

 Now we consider the closed string transition amplitude between a pair
of permutation Ishibashi states with the permutations $\Om'$ and $\Om$.
It is given by the following expression
\ber
<<I_{\bh'},\Om'\et,B|(-1)^{g(\Om',\Om)}q^{L[0]-{c\ov
24}}u^{J[0]}|I_{\bh},\Om,\et,B>>=\nmb
\dlt(\bh-\bh')\dlt(\Om'\bh'-\bh')\dlt(\Om\bh-\bh)
\nmb
\sum_{(\bp,\bp^{*})\in\Gm_{\bh}}(-1)^{g(\Om',\Om)}|c_{\bp,\bp^{*}}|^{2}
\dlt(\Om'\Om^{-1}\bp-\bp)
\dlt(\Om'\Om^{-1}\bp^{*}-\bp^{*})
\nmb
<<\bp,\bp^{*},\Om',\et|(-1)^{g(\Om',\Om)}q^{L[0]-{c\ov
24}}u^{J[0]}
|\bp,\bp^{*},\Om,\et,B>>.
\label{3.IBamp}
\enr
Due to the insertion $(-1)^{g(\Om',\Om)}$ the amplitude is calculating 
according to the ghost number of the intermediate closed string states and 
the ghost number operator $g(\Om',\Om)$ depends on the permutation matrices.  
To simplify the calculation we put here the number $N$
of permutation groups to be equal 1 (and hence $\mu_{1}=...\mu_{r}=\mu$).
Due to the factor $\dlt(\Om'\Om^{-1}\bp-\bp)\dlt(\Om'\Om^{-1}\bp^{*}-\bp^{*})$
the summation is restricted to the subspace of $\Gm_{\bh}$ which
is invariant with respect to the permutation $\Om'\Om^{-1}$. 
It allows us to
write
\ber
<<\bp,\bp^{*},\Om',\et|(-1)^{g(\Om',\Om)}q^{L[0]-{c\ov
24}}u^{J[0]}
|\bp,\bp^{*},\Om,\et,B>>=
\nmb
q^{{1\ov 2}(|\Xi|_{1}(2p^{*}_{1}p_{1}+p_{1}+{p^{*}_{1}\ov\mu})+
...|\Xi|_{\nu(\Xi)}(2p^{*}_{\nu(\Xi)}p_{\nu(\Xi)}+p_{\nu(\Xi)}
+{p^{*}_{\nu(\Xi)}\ov\mu})-{c\ov24}}
\nmb
u^{(|\Xi|_{1}({p^{*}_{1}\ov\mu}-p_{1})+
...
+|\Xi|_{\nu(\Xi)}
({p^{*}_{\nu(\Xi)}\ov\mu}-p_{\nu(\Xi)})}(oscillator \ contribution),
\label{3.IBamp1}
\enr
where $|\Xi|_{i}$ is the length of $i_{th}$ cycle of the permutation
$\Xi\equiv \Om'\Om^{-1}$ and $\nu(\Xi)$ is the number of cycles of the 
permutation.

 The oscillator contribution calculation is useful to carry out 
for the bosons and fermions separately.  The bosonic contribution
can be calculated as follows. First of all we have from (\ref{3.BI})
\ber
\prod_{a,c}
<\bp,\bp^{*},-(\Om')^{T}\bp-\bd,-(\Om')^{T}\bp^{*}-\bd^{*}|
\prod_{n=1}\exp(-{1\ov n}X^{*}_{a}[n]\bar{X}_{\Om'(a)}[n])
\nmb
\prod_{m=1}\exp(-{q^{m}\ov m}
X_{c}[-m]\bar{X}^{*}_{\Om(c)}[-m]) 
|\bp,\bp^{*},-\Om^{T}\bp-\bd,-\Om^{T}\bp^{*}-\bd^{*}>=
\nmb 
\prod_{n=1}\sum_{k_{1},...,k_{r}=0}\sum_{l_{1},...,l_{r}=0}
{1\ov k_{1}!...k_{r}!l_{1}!...l_{r}!}
{q^{n(l_{1}+...+l_{r})}\ov n^{k_{1}+...+l_{r}}}
\nmb
<\bp,\bp^{*},-(\Om')^{T}\bp-\bd,-(\Om')^{T}\bp^{*}-\bd^{*}|
(X^{*}_{1}[n])^{k_{1}}(\bar{X}_{\Om'(1)}[n])^{k_{1}}...
(X^{*}_{r}[n])^{k_{r}}(\bar{X}_{\Om'(r)}[n])^{k_{r}}
\nmb
(X_{1}[-n])^{l_{1}}(\bar{X}^{*}_{\Om(1)}[-n])^{l_{1}}...
(X_{r}[-n])^{l_{r}}(\bar{X}^{*}_{\Om(r)}[-n])^{l_{r}}
|\bp,\bp^{*},-\Om^{T}\bp-\bd,-\Om^{T}\bp^{*}-\bd^{*}>=
\nmb
\prod_{n=1}(1-q^{n|\Xi|_{1}})^{-1}...
(1-q^{n|\Xi|_{\nu(\Xi)}})^{-1}.
\label{3.IBamp2}
\enr
By the similar reasons
\ber
\prod_{a,c}
<\bp,\bp^{*},-(\Om')^{T}\bp-\bd,-(\Om')^{T}\bp^{*}-\bd^{*}|
\prod_{n=1}\exp(-{1\ov n}X_{a}[n]\bar{X}^{*}_{\Om'(a)}[n])
\nmb
\prod_{m=1}\exp(-{q^{m}\ov m}(X^{*}_{c}[-m]\bar{X}_{\Om(c)}[-m]))
|\bp,\bp^{*},-\Om^{T}\bp-\bd,-\Om^{T}\bp^{*}-\bd^{*}>=
\nmb
\prod_{n=1}(1-q^{n|\Xi|_{1}})^{-1}...
(1-q^{n|\Xi|_{\nu(\Xi)}})^{-1}.
\label{3.IBamp3}
\enr

 The first part of the fermionic contribution is given by
\ber
\prod_{b,d}
<\bp,\bp^{*},-(\Om')^{T}\bp-\bd,-(\Om')^{T}\bp^{*}-\bd^{*}|
\prod_{r=1/2}(1-\im\et\psi^{*}_{b}[r]\bar{\psi}_{\Om^{-1}(b)}[r])
\nmb
\prod_{s=1/2}(1+\im\et u^{-1}q^{s}\psi_{d}[-s]\bar{\psi}^{*}_{\Om(d)}[-s])
|\bp,\bp^{*},-\Om^{T}\bp-\bd,-\Om^{T}\bp^{*}-\bd^{*}>=
\nmb
\prod_{s=1/2}\sum_{k=0}^{r}Tr(\Xi)|_{\wedge^{k}V}u^{-k}q^{ks},
\label{3.IBamp4}
\enr
where $Tr(\Xi)|_{\wedge^{k}V}$ means the trace of matrix $\Xi=\Om'\Om^{-1}$
acting by permutation components in the space $\wedge^{k}V$ of the 
$r$-dimensional real vector space $V$. One can see that the last 
expression can be rewritten similar to (\ref{3.IBamp2})
\ber
\prod_{b,d}
<\bp,\bp^{*},-(\Om')^{T}\bp-\bd,-(\Om')^{T}\bp^{*}-\bd^{*}|
\prod_{r=1/2}(1-\im\et(\psi^{*}_{b}[r]\bar{\psi}_{(\Om')^{-1}(b)}[r])
\nmb
\prod_{s=1/2}(1+\im\et u^{-1}q^{s}\psi_{d}[-s]\bar{\psi}^{*}_{\Om(d)}[-s])
|\bp,\bp^{*},-\Om^{T}\bp-\bd,-\Om^{T}\bp^{*}-\bd^{*}>=
\nmb
\prod_{s=1/2}(1-(-1)^{|\Xi|_{1}}u^{-|\Xi|_{1}}
q^{s|\Xi|_{1}})...
(1-(-1)^{|\Xi|_{\nu(\Xi)}}u^{-|\Xi|_{\nu(\Xi)}}
q^{s|\Xi|_{\nu(\Xi)}}). 
\label{3.IBamp5}
\enr
Analogously
\ber
\prod_{a,c}
<\bp,\bp^{*},-(\Om')^{T}\bp-\bd,-(\Om')^{T}\bp^{*}-\bd^{*}|
\prod_{r=1/2}(1-\im\et\psi^{*}_{a}[r]\bar{\psi}_{(\Om')^{-1}(a)}[r])
\nmb
\prod_{s=1/2}(1+\im\et u^{-1}q^{s}\psi_{c}[-s]\bar{\psi}^{*}_{\Om(c)}[-s])
|\bp,\bp^{*},-\Om^{T}\bp-\bd,-\Om^{T}\bp^{*}-\bd^{*}>=
\nmb
\prod_{s=1/2}(1-(-1)^{|\Xi|_{1}}u^{|\Xi|_{1}}
q^{s|\Xi|_{1}})...
(1-(-1)^{|\Xi|_{\nu(\Xi)}}u^{|\Xi|_{\nu(\Xi)}}
q^{s|\Xi|_{\nu(\Xi)}}). 
\label{3.IBamp6}
\enr
Collecting the results we obtain
\ber
<<I_{\bh'},\Om',\et,B|(-1)^{g(\Om',\Om)}q^{L[0]-{c\ov24}}u^{J[0]}
|I_{\bh},\Om,\et,B>>=
\nmb
\dlt(\bh-\bh')\dlt(\Om'\bh-\bh)\dlt(\Om\bh-\bh)|c_{0,\bh}|^{2}
\sum_{(\bp,\bp^{*})\in\Gm_{\bh}}(-1)^{g(\Om',\Om)}
\dlt(\Xi\bp-\bp)
\dlt(\Xi\bp^{*}-\bp^{*})
\nmb
q^{{1\ov 2}(\sum_{i=1}^{\nu(\Xi)}|\Xi|_{i}(2p^{*}_{i}p_{i}+p_{i}+{p^{*}_{i}\ov\mu}))
-{c\ov24}}
u^{\sum_{i=1}^{\nu(\Xi)}|\Xi|_{i}({p^{*}_{i}\ov\mu}-p_{i})}
\nmb
\prod_{i=1}^{\nu(\Xi)}\prod_{n=1}(1-(-1)^{|\Xi|_{i}}u^{-|\Xi|_{i}}
q^{(n-1/2)|\Xi|_{i}})
(1-(-1)^{|\Xi|_{i}}u^{|\Xi|_{i}}
q^{(n-1/2)|\Xi|_{i}})
(1-q^{n|\Xi|_{i}})^{-2}.
\label{3.IBamp7}
\enr 
 
 The transition amplitude between $|I_{\bh'},\Om',-\et,B>>$ and
$|I_{\bh},\Om,\et,B>>$ is given by the similar expression. Indeed,
the change $\et\rightarrow -\et$ affects only on the fermionic contribution
(\ref{3.IBamp4})-(\ref{3.IBamp6}) so that
\ber
<<I_{\bh'},\Om',-\et,B|(-1)^{g(\Om',\Om)}q^{L[0]-{c\ov 24}}u^{J[0]}
|I_{\bh},\Om,\et,B>>=
\nmb
\dlt(\bh-\bh')\dlt(\Om'\bh-\bh)\dlt(\Om\bh-\bh)|c_{0,\bh}|^{2}
\sum_{(\bp,\bp^{*})\in\Gm_{\bh}}(-1)^{g(\Om',\Om)}
\dlt(\Xi\bp-\bp)
\dlt(\Xi\bp^{*}-\bp^{*})
\nmb
q^{{1\ov 2}(\sum_{i=1}^{\nu(\Xi)}|\Xi|_{i}(2p^{*}_{i}p_{i}+p_{i}+{p^{*}_{i}\ov\mu}))
-{c\ov24}}
u^{\sum_{i=1}^{\nu(\Xi)}|\Xi|_{i}({p^{*}_{i}\ov\mu}-p_{i})}
\nmb
\prod_{i=1}^{\nu(\Xi)}\prod_{n=1}(1-u^{-|\Xi|_{i}}
q^{(n-1/2)|\Xi|_{i}})
(1-u^{|\Xi|_{i}}
q^{(n-1/2)|\Xi|_{i}})
(1-q^{n|\Xi|_{i}})^{-2}.
\label{3.IBampet}
\enr 
  
 Now one can fix the dependence of the ghost number operator on the permutation
matrices. Taking into account the representation (\ref{1.char}) we find
that the amplitude is given by the product of minimal model
characters if 
\ber
g(\Om',\Om)=\sum_{i=1}^{\nu(\Xi)}g_{i}.
\label{3.ghostoper}
\enr
Thus the ghost number receives the contribution $g_{i}$ from
$i_{th}$ invariant subspace of $\Gm_{\bh}$. In other words we consider the
space of intermediate closed string states as a product of minimal model
butterfly resolutions (\ref{1.but}) in amount of the number $\nu(\Xi)$.
Hence the amplitude is given by the following product of minimal model
characters 
\ber
<<I_{\bh'},\Om',\et,B|(-1)^{g(\Om',\Om)}q^{L[0]-{c\ov24}}u^{J[0]}|I_{\bh},\Om,\et,B>>=
\nmb
\dlt(\bh-\bh')\dlt(\Om'\bh-\bh)\dlt(\Om\bh-\bh)|c_{0,\bh}|^{2}
\prod_{i=1}^{\nu(\Xi)}
\exp(-\im\pi(1-|\Xi|_{i}){h_{i}\ov\mu})
\chi_{h_{i}}(\tau,\ups+{1-|\Xi|_{i}\ov 2}),
\label{4.IBamp8}
\enr
where $f$ is fermion number operator and we have used the relation
\ber
Tr_{M_{h_{i}}}((-1)^{(1-|\Xi|_{i})f}
q^{(L_{i}[0]-{c_{i}\ov 24})}u^{J_{i}[0]})=
\exp(-\im\pi(1-|\Xi|_{i}){h_{i}\ov\mu})
\chi_{h_{i}}(\tau,\ups+{1-|\Xi|_{i}\ov 2}).
\label{4.tldchar}
\enr
Analogously
\ber
<<I_{\bh'},\Om',-\et,B|(-1)^{g(\Om',\Om)}q^{L[0]-{c\ov24}}u^{J[0]}
|I_{\bh},\Om,\et,B>>=
\nmb
\dlt(\bh-\bh')\dlt(\Om'\bh-\bh)\dlt(\Om\bh-\bh)|c_{0,\bh}|^{2}
\prod_{i=1}^{\nu(\Xi)}
\exp(-\im\pi{h_{i}\ov\mu})
\chi_{h_{i}}(\tau,\ups+{1\ov 2}),
\label{4.IBampet8}
\enr

 It was mentioned in Section 2 that the irreducible representations
are generated by the spectral flow action. Hence for an arbitrary module 
$M_{\bh,\bt}$, $(\bh,\bt)\in \Dl$, the Ishibashi state is generated by the 
action of spectral flow operators on the Ishibashi state (\ref{4.Isuper}). 
It is easy to check that the state
\ber
|I_{\bh,\bt},\Om,\et,B>>=
\prod_{i}U_{i}^{t_{i}}\bar{U}_{i}^{-t_{i}}
|I_{\bh},\Om,\et,B>>,
\label{4.IBht}
\enr
satisfy B-type boundary conditions if
\ber
\Om\bt-\bt=0.
\label{4.sftrest}
\enr

 One has to take into account however that the right-moving
space of states of the model is governed by dual butterfly resolutions 
(twisted by right-moving spectral flow operators). The representative of 
chiral primary field from the dual resolution is 
\ber
U^{h}G^{+}[{1\ov 2}-h]...G^{+}[-{1\ov 2}]|-{1\ov\mu},-1-h>\sim 
|-{1+h\ov\mu},-1>.
\label{4.dualchw}
\enr
Thus the highest-weight vectors of the model are given by
the products of the following minimal model states
\ber
|p_{i}={t_{i}\ov\mu_{i}},p^{*}_{i}=h_{i}-t_{i},
\bar{p}_{i}=-{1+h_{i}-t_{i}-l\ov\mu_{i}},\bar{p}^{*}_{i}=-1-t_{i}-l>.
\label{4.dualchw1}
\enr
Therefore the Ishibashi states (\ref{4.IBht}) have nontrivial overlap with
the states (\ref{4.dualchw1}) if in addition to (\ref{4.sftrest})
we have 
\ber
h_{\Om^{-1}(i)}=h_{i}, 
\nmb
h_{i}-2t_{i}-l=0 \ mod\ \mu_{i}.
\label{4.chrgconj}
\enr

 It is easy to see from (\ref{1.3}) that this state satisfy the
boundary conditions (\ref{3.Bantz}), (\ref{3.BX}). Hence
(\ref{3.BD}) is fulfilled. It is also $BRST$ closed because
the spectral flow commutes with screening charges. 

 The transition amplitude between such states is spectral flow 
twist of the amplitudes (\ref{4.IBamp8}), (\ref{4.IBampet8}).
\ber
<<I_{\bh',\bt'},\Om',\et,B|
(-1)^{g(\Om',\Om)}q^{L[0]-{c\ov24}}u^{J[0]}|I_{\bh,\bt},\Om,\et,B>>=
\nmb
\dlt(\bh-\bh')\dlt(\Om'\bh-\bh)\dlt(\Om\bh-\bh)
\dlt^{(\bmu)}(\bt-\bt')
\nmb
|c_{0,\bh}|^{2}
\prod_{i=1}^{\nu(\Xi)}
\exp(-\im\pi(1-|\Xi|_{i}){h_{i}-2t_{i}\ov\mu})
\chi_{h_{i},t_{i}}(\tau,\ups+{1-|\Xi|_{i}\ov 2}).
\label{4.IBamp9}
\enr
\ber
<<I_{\bh',\bt'},\Om',-\et,B|
(-1)^{g(\Om',\Om)}q^{L[0]-{c\ov24}}u^{J[0]}|I_{\bh,\bt},\Om,\et,B>>=
\nmb
\dlt(\bh-\bh')\dlt(\Om'\bh-\bh)\dlt(\Om\bh-\bh)
\dlt^{(\bmu)}(\bt-\bt')
\nmb
|c_{0,\bh}|^{2}
\prod_{i=1}^{\nu(\Xi)}
\exp(-\im\pi{h_{i}-2t_{i}\ov\mu})
\chi_{h_{i},t_{i}}(\tau,\ups+{1\ov 2}).
\label{4.IBampet9}
\enr

\vskip 10pt
\leftline {\bf 4.2. $A$-type permutation Ishibashi states.}

 Let us consider free-field representation for $A$-type
Ishibashi states. It is obvious that $A$-type Ishibashi states are given
by superpositions like (\ref{4.Isuper}).

 Similar to the $B$-type case one can conclude that matrix $\Ups^{T}$ is proportional
to the element of the permutation group $\aleph_{r_{1}...r_{N}}$. More precisely 
\ber
\Ups=\mu_{1}\Om_{1}\otimes...\otimes \mu_{N} \Om_{N},
\
\Ups^{*}={1\ov\mu_{1}}\Om_{1}\otimes...\otimes {1\ov\mu_{N}} \Om_{N}
\label{4.Amatr}
\enr
where $\Om_{i}\in \aleph_{r_{i}}$, $i=1,...,N$.

 The boundary conditions (\ref{3.Ator}) take the form which is mirror
to (\ref{4.Btorr})
\ber
(\sgm_{i}[s]-\im\et \Om_{ij}\bar{\sgm}_{j}[-s])|A>>=0, \nmb
(\gm_{i}[s]+\im\et \Om_{ij}\bar{\gm}_{j}[-s])|A>>=0,
\nmb
(\bar{R}_{j}[-n]+\Om_{ij}R_{i}[n]+
\sqrt{{2\ov\mu_{j}}}\dlt_{n,0})|A>>=0,
\nmb
(\bar{\theta}_{j}[-n]-\Om_{ij}\theta_{i}[n])|A>>=0.
\label{4.Atorr}
\enr

 The $BRST$ condition for $A$-type states is slightly different
from $B$-type case. The reason is that the application of one of the
left-moving $BRST$ charges, say $Q^{+}_{i}$ to $A$-type state
gives according to (\ref{3.Aantz}) and (\ref{3.AX}) the
right-moving $BRST$ charge $\bar{Q}^{-}_{G^{-1}(i)}$ multiplied by
$\mu_{i}$ as opposed to the $B$-type case.
In fact we are free to rescale arbitrary the right-moving $BRST$ charges
because it does not change the cohomology of the complex in the
right-moving sector and the cohomology of the total complex
(\ref{4.complex}).
Hence we define the right-moving $BRST$ charges in such a way to
cancel this effect
\ber
\bar{S}^{+}_{i}(\bar{z})=
{\im\et\ov\mu_{i}}{\bf s}_{i}\bar{\psi}^{*}\exp({\bf s}_{i}\bar{X}^{*})(\bar{z}), \nmb
\bar{S}^{-}_{i}(\bar{z})=
\im\et\mu_{i}{\bf s}^{*}_{i}\bar{\psi}\exp(\mu_{i}{\bf s}^{*}_{i}\bar{X})(\bar{z}), \nmb
\bar{Q}^{\pm}_{i}=\oint d\bar{z} \bar{S}^{\pm}_{i}(\bar{z}),
\label{4.abarscr}
\enr

 As a result $BRST$ invariant
$A$-type Ishibashi state $|I_{\bh},\Om,\et,A>>$ is given similar to
(\ref{4.Isuper}), (\ref{3.c}) with the restriction
$\dlt(\Om\bh-\bh)$ and similar to $B$-type case the phase
of coefficient $c_{0,\bh}$ is arbitrary also. 

 $A$-type version of the transition amplitude (\ref{4.IBamp8})
can be calculated similar to $B$-type case so the result is given by
\ber
<<I_{\bh'},\Om',\et,A|(-1)^{g(\Om',\Om)}q^{L[0]-{c\ov24}}u^{J[0]}|I_{\bh},\Om,\et,A>>=
\nmb
\dlt(\bh-\bh')\dlt(\Om'\bh-\bh)\dlt(\Om\bh-\bh)
|c_{0,\bh}|^{2}
\nmb
\prod_{i=1}^{\nu(\Xi)}
\exp(-\im\pi(1-|\Xi|_{i}){h_{i}\ov\mu})
\chi_{h_{i}}(\tau,\ups+{1-|\Xi|_{i}\ov 2}).,
\label{4.Iaamp}
\enr
where $\Xi=\Om'\Om^{-1}$ and we have puted for simplicity the number $N$
of permutation groups to be equal 1. 

 For an arbitrary module 
$M_{\bh,\bt}$, $(\bh,\bt)\in \Dl$, the A-type Ishibashi state is generated by 
the action of spectral flow operators. 
It is easy to check that the state
\ber
|I_{\bh,\bt},\Om,\et,A>>=
\prod_{i}U_{i}^{t_{i}}\bar{U}_{i}^{t_{i}}
|I_{\bh},\Om,\et,A>>,
\label{4.IAht}
\enr
satisfy A-type boundary condition if the spectral flow parameter $\bt$ satisfy
(\ref{4.sftrest}).Though the right-moving
space of states of the model is governed by dual butterfly resolutions 
(twisted by right-moving spectral flow operators) the only restrictions
on $\bh,\bt$ are
\ber
\Om\bh=\bh, \
\Om\bt=\bt.
\label{4.sfrestA}
\enr

 Corresponding transition amplitude is given similar to B-type case.
\ber
<<I_{\bh',\bt'},\Om',\et,A|(-1)^{g(\Om',\Om)}q^{L[0]-{c\ov24}}u^{J[0]}
|I_{\bh,\bt},\Om,\et,A>>=
\nmb
\dlt(\bh-\bh')\dlt(\Om'\bh-\bh)\dlt(\Om\bh-\bh)
\dlt^{\bmu}(\bt-\bt')
|c_{0,\bh}|^{2}
\nmb
\prod_{i=1}^{\nu(\Xi)}
\exp(-\im\pi(1-|\Xi|_{i}){h_{i}-2t_{i}\ov\mu})
\chi_{h_{i},t_{i}}(\tau,\ups+{1-|\Xi|_{i}\ov 2}).
\label{4.Iaamp}
\enr
\ber
<<I_{\bh',\bt'},\Om',-\et,A|(-1)^{g(\Om',\Om)}q^{L[0]-{c\ov24}}u^{J[0]}
|I_{\bh,\bt},\Om,\et,A>>=
\nmb
\dlt(\bh-\bh')\dlt(\Om'\bh-\bh)\dlt(\Om\bh-\bh)
\dlt^{\bmu}(\bt-\bt')
|c_{0,\bh}|^{2}
\nmb
\prod_{i=1}^{\nu(\Xi)}
\exp(-\im\pi{h_{i}-2t_{i}\ov\mu})
\chi_{h_{i},t_{i}}(\tau,\ups+{1\ov 2}).
\label{4.Iaampet}
\enr
 Thus the expressions (\ref{4.IBamp9}), (\ref{4.Iaamp})
reproduce correctly (with the correct fermionic contribution) the 
corresponding results from ~\cite{Reper}. It allows to use the
Cardy's constraint solution found for permutation branes in ~\cite{Reper}
to construct the free-field representation of permutation branes.  


\vskip 10pt
\centerline{\bf 5. Free-field representation of permutation branes in}
\centerline{\bf Gepner model.}
\vskip 10pt
\leftline {\bf 5.1. A-type boundary states in Calabi-Yau extension.}

 It has already been noticed that the product of minimal models can not be
applied straightforward to describe in the bulk the string theory on
CY manifold. Instead, the so called simple current orbifold whose
partition function is diagonal modular invariant partition function with
respect to orbit characters (\ref{2.orbchi})
describes. The extension of this technique to the conformal field theory
with a boundary has been developed in ~\cite{ReS},~\cite{Reper}, ~\cite{FSW},
~\cite{BRS}, ~\cite{GJ}.

 As we have seen $BRST$ invariance fixes the free-field permutation Ishibashi 
states up to the arbitrary constant $c_{\bh,\bt}$. Hence our problem is to
apply the (simple current) orbifold construction and Cardy's constraint to 
the superposition of free-field permutation Ishibashi states with arbitrary 
coefficients $c_{\bh,\bt}$. Fortunately the Cardy's constraint for the 
perturbation branes has been found by Recknagel ~\cite{Reper}.
Hence it is sufficient only to quote his solution.

 Thus the free-field realization of permutation A-type branes can be given as 
follows. We start first from the spectral flow invariant permutation boundary 
states
\ber
|[\bLm,\blm],\Om,\et,A>>={\al\ov\kp^{2}}
\sum_{(\bh,\bt)\in \tld{\Dl}}\dlt(\Om\bh-\bh)\dlt(\Om\bt-\bt)
\nmb
W^{\bh,\bt}_{\bLm,\blm,\Om}
\sum_{m,n=0}^{\kp-1}\exp{(\im 2\pi nJ[0])}
U^{m\bv}\bar{U}^{m\bv}|I_{\bh,\bt},\Om,\et,A>>.
\label{5.DA}
\enr
They are labeled by the spectral flow orbit classes $[\bLm,\blm]$ 
of the vectors $(\bLm,\blm)\in \Dl$.
The coefficients $W^{\bh,\bt}_{\bLm,\blm,\Om}$ which solve the Cardy's constraint
are given by ~\cite{Reper}
\ber
W^{\bh,\bt}_{\bLm,\blm,\Om}=
\prod_{a=1}^{\nu(\Om)}S_{(\Lm_{a},\lm_{a})(h_{a},t_{a})}
(S_{(0,0),(h_{a},t_{a})})^{-{1\ov 2}|\Om|_{a}},
\nmb
S_{(\Lm_{a},\lm_{a})(h_{a},t_{a})}=S_{\Lm_{a},h_{a}}
\exp(\im\pi{(h_{a}-2t_{a})(\Lm_{a}-2\lm_{a})\ov\mu}),
\nmb
S_{\Lm_{a},h_{a}}={\sqrt{2}\ov\mu}\sin(\pi{(h_{a}+1)(\Lm_{a}+1)\ov\mu}).
\label{5.ReCard}
\enr
The summation over $n$ makes $J[0]$-projection, while summation over $m$ introduce
spectral flow twisted sectors.
This state depends only on the spectral flow orbit class.
Moreover, the $J[0]$ integer charge restriction of the orbits $[\bLm,\blm]$ 
is necessary for the self-consistency of the expression (\ref{5.DA}).
$\al$ is the normalization constant.

 Now we apply the internal automorphism group of Gepner model to construct additional
boundary states. Namely one can use the operator
$\exp{(-\im2\pi\sum_{i}\phi_{i}J_{i}[0])}\in U(1)^{r}$ to generate new
boundary states. Let us consider the properties of the state
\ber
|[\bLm,\blm],\Om,\et,A>>_{\bphi}
\equiv\exp{(-\im2\pi\sum_{i}\phi_{i}J_{i}[0])}|[\bLm,\blm],\Om,\et,A>>.
\label{5.DAphi}
\enr
It satisfies the conditions similar to (\ref{3.AD}) except the
relations for fermionic fields
\ber
(G^{\pm}[r]+\im\et\sum_{i}\exp{(\pm\im2\pi\phi_{\Om(i)})}\bar{G}^{\mp}_{\Om(i)}[-r])
|[\bLm,\blm],\Om,\et,A>>_{\bphi}=0,
\nmb
(\psi_{i}^{*}[r]-\im\et\mu_{i}\exp{(\im2\pi\phi_{\Om(i)})}\bar{\psi}_{\Om(i)}[-r])
|[\bLm,\blm],\Om,\et,A>>_{\bphi}=0,
\nmb
(\psi_{i}[r]-\im{\et\ov\mu_{i}}\exp{(-\im2\pi\phi_{\Om(i)})}\bar{\psi}^{*}_{\Om(i)}[-r])
|[\bLm,\blm],\Om,\et,A>>_{\bphi}=0.
\label{5.Aphi}
\enr
This state does not invariant with respect to diagonal N=2 Virasoro algebra unless
\ber
\phi_{i}\in Z, \ i=1,...,r.
\label{5.Aphi1}
\enr
Hence the group $U(1)^{r}$ reduces to $Z^{r}$. It is worth to note that one can ignore
the case when all $\phi_{i}$ are half-integer because it can be canceled by the $\et\rightarrow -\et$
redefinition.
It is easy to see directly what kind of states we obtain by this way.
\ber
|[\bLm,\blm]\Om,\et,A>>_{\bphi}=
\nmb
{\al\ov \kp^{2}}
\sum_{(\bh,\bt)\in \Dl_{\Om}}W^{\bh,\bt}_{[{\bf\Lm},{\bf\lm}]}
\sum_{m,n=0}^{\kp-1}\exp{(\im 2\pi nJ[0])}\exp{(-\im2\pi m\sum_{i}\phi_{i}{c_{i}\ov3})}U^{m\bv}\bar{U}^{m\bv}
\nmb
\exp{(-\im2\pi\sum_{i}\phi_{i}{h_{i}-2t_{i}\ov\mu_{i}})}|I_{\bh,\bt},\Om,\et,A>>=
\nmb
{\al\ov \kp^{2}}
\sum_{(\bh,\bt)\in \Dl_{\Om}}\prod_{e=1}^{\nu(\Om)}S_{\Lm_{e},h_{e}}
(S_{0,h_{e}})^{-{|\Om|_{e}\ov 2}}
\exp{(\im\pi{(\Lm_{e}-2\lm_{e})(h_{e}-2t_{e})\ov\mu})}
\nmb
\exp{(-\im\pi{(h_{e}-2t_{e})\ov\mu}\sum_{a=1}^{|\Om|_{e}}2\phi_{e+a})}
\exp{(\im4\pi m\sum_{a=1}^{|\Om|_{e}}{\phi_{e+a}\ov\mu})}
\nmb
\sum_{m,n=0}^{\kp-1}\exp{(\im 2\pi nJ[0])}
U^{m\bv}\bar{U}^{m\bv}
|I_{\bh,\bt},\Om,\et,A>>.
\label{5.DAphi1}
\enr
It allows to parameterize the boundary states by
\ber
|[\bLm,\blm],\Om,\et,A>>=\exp{(-\im 2\pi\sum_{i}\lm_{i}J_{i}[0])}
|[\bLm,0],\Om,\et,A>>
\label{5.Jparam}
\enr
so that the different boundary states are labeled by different values of
$|\lm|_{e}=\sum_{a=1}^{|\Om|_{e}}\lm_{e+a}$, $e=1,...,\nu(\Om)$ and 
spectral flow invariant boundary states are recovered when
\ber
{2\ov \mu}\sum_{e=1}^{\nu(\Om)}|\lm|_{e}\in Z.
\label{5.Asfinv}
\enr

\vskip 10pt
\leftline {\bf 5.2. B-type boundary states in Calabi-Yau extension.}

 Let us denote by $\Dl_{\Om}$ the subset of
$\Dl$ satisfying (\ref{4.sftrest}), (\ref{4.chrgconj}). Then
for an arbitrary pair of vectors $(\bLm,\blm)\in \Dl_{CY}$ the
free-filed realization of spectral flow invariant $B$-type boundary state is given by
\ber
|[\bLm,\blm],\Om,\et,B>>={\al\ov\kp^{2}}
\sum_{(\bh,\bt)\in {\Dl_{\Om}}}W^{\bh,\bt}_{[\bLm,\blm],\Om}
\sum_{m,n=0}^{\kp-1}\exp{(\im 2\pi nJ[0])}
U^{m\bv}\bar{U}^{-m\bv}|I_{\bh,\bt},\Om,\et,B>>,
\label{5.DB}
\enr
where the coefficients $W^{\bh,\bt}_{\bLm,\blm,\Om}$ are given by
(\ref{5.ReCard}).
One can check that this state depends only on the spectral flow orbit
class $[\bLm,\blm]$ of vectors $(\bLm,\blm)$. It is also obvious that
$[\bLm,\blm]$ has to be restricted to the set of $J[0]$ integer charges by the 
reasons similar to the $A$-type case.

 The other boundary states are generated by internal automorphism group of 
Gepner model similar to the A-type case.  
Namely the state
\ber
|[\bLm,\blm],\Om,\et,B>>_{\bphi}
\equiv\exp{(-\im2\pi\sum_{i}\phi_{i}J_{i}[0])}|[\bLm,\blm],\Om,\et,B>>
\label{5.DBphi}
\enr
satisfies the conditions similar to (\ref{5.Aphi})
and does not invariant with respect to diagonal N=2 Virasoro algebra unless
\ber
\phi_{i}\in Z, \ i=1,...,r.
\label{5.Bphi1}
\enr
Hence the group $U(1)^{r}$ reduces to $Z^{r}$
and one can parameterize the boundary states by
\ber
|[\bLm,\blm],\Om,\et,B>>=\exp{(-\im 2\pi\sum_{i}\lm_{i}J_{i}[0])}
|[\bLm,0],\Om,\et,B>>
\label{5.BJparam}
\enr
so that the different boundary states are labeled by different values of
$|\lm|_{e}=\sum_{a=1}^{|\Om|_{e}}\lm_{e+a}$, $e=1,...,\nu(\Om)$.

 In conclusion of this section we would like to make the following remarks.
First of all we note that our free-field construction allows to interpretate
of A/B-type gluing conditions (\ref{3.AD}), (\ref{3.BD}) geometrically. Indeed,
in terms of the free-fields B-type gluing conditions for example are given by (\ref{4.Btorr}).
Thus $\pm 1$ eigne-values of the permutation matrix $\Om$ can be interpreted as labeling
Newman and Dirichlet boundary conditions. While the complex eigne-values realize mixed 
boundary conditions ~\cite{OOY}. This result seems to contradict to the calculation 
of D-brane charges performed in
~\cite{BrGa} and ~\cite{ERR}. It has been found there that D0-branes correspond to 
transposition matrices permuting
only one pair of minimal models. It follows from (\ref{4.Btorr}) that in this case we have 
only one Dirichlet condition and the corresponding free-field boundary state gives 
codimension one D-brane. We do not know at the moment how to resolve or explain the 
contradiction. Perhaps more profound 
geometric investigation of the open string spectrum in terms of chiral de Rham complex
has to be performed. But we postpone it for the next publication.

 As the next remark we note that the free-field
representations of permutation boundary states are
determined modulo $BRST$-exact states satisfying $A$ or $B$-type boundary conditions.
We interpret this ambiguity in the free-field representation as a result of adding
brane-antibrane pairs annihilating under the tachyon condensation process
~\cite{Sen}. In this context the free-field representations of boundary states can be
considered as the superpositions of branes flowing under the tachyon condensation to
nontrivial boundary states in Gepner models.
  It is also important to note that the automorphisms (\ref{1.reflect}) give
different free-field representations of boundary states because the corresponding
butterfly resolutions are not invariant with respect to these automorphisms.
However their cohomology are invariant. Hence these different representations have to 
be identified. Thus the free-field boundary states construction have to be 
considered in derived category sense ~\cite{DCat}.

\vskip 10pt
\leftline {\bf 5.3. Free-field representation of permutation boundary states in 
Gepner models.}

 It is completely clear from (\ref{2.gschar}) and (\ref{2.ZG})
haw to incorporate in our construction the space-time
degrees of freedom to obtain the free-field construction of
permutation branes in Calabi-Yau extension to the case of Gepner models.
It is straightforward (see for example ~\cite{ReS}, ~\cite{BDLR},~\cite{FSW},
~\cite{SP}) and we shall not represent the details here.

\vskip 10pt
\centerline{\bf Acknowledgments}

 This work was supported in part by grants RBRF-04-02-16027,
SS 2044-2003.2003.2, INTAS-OPEN-03-51-3350, .

\end{document}